\documentclass{aa}

\usepackage{graphicx}
\usepackage{txfonts}

\usepackage{amsmath}	
\usepackage{amssymb}	
\usepackage{mathtools}
\usepackage{multicol} 
\usepackage{multirow} 
\usepackage{threeparttable}
\usepackage{longtable}
\usepackage{tabto}
\usepackage{adjustbox}

\usepackage{color}
\usepackage{natbib,twoopt}
\usepackage[hyphenbreaks]{breakurl}
\usepackage[breaklinks]{hyperref}      
\bibpunct{(}{)}{;}{a}{}{,}             
\definecolor{cobalt}{rgb}{0.06, 0.2, 0.65}
\hypersetup{
  colorlinks,
  citecolor=blue,
  linkcolor=blue,
  urlcolor=blue,
}
\makeatletter
  \newcommandtwoopt{\citeads}[3][][]{\href{http://adsabs.harvard.edu/abs/#3}%
    {\def\hyper@linkstart##1##2{}%
     \let\hyper@linkend\@empty\citealp[#1][#2]{#3}}}
  \newcommandtwoopt{\citepads}[3][][]{\href{http://adsabs.harvard.edu/abs/#3}%
    {\def\hyper@linkstart##1##2{}%
     \let\hyper@linkend\@empty\citep[#1][#2]{#3}}}
  \newcommandtwoopt{\citetads}[3][][]{\href{http://adsabs.harvard.edu/abs/#3}%
    {\def\hyper@linkstart##1##2{}%
     \let\hyper@linkend\@empty\citet[#1][#2]{#3}}}
  \newcommandtwoopt{\citeyearads}[3][][]%
    {\href{http://adsabs.harvard.edu/abs/#3}
    {\def\hyper@linkstart##1##2{}%
     \let\hyper@linkend\@empty\citeyear[#1][#2]{#3}}}
\makeatother

\newcommand{\Lsun}{L$_{\odot}$}
\newcommand{\Msun}{M$_{\odot}$}
\newcommand{\Rsun}{R$_{\odot}$}
\newcommand{\Msunyr}{M$_{\odot}$~yr$^{-1}$}

\definecolor{smalt(darkpowderblue)}{rgb}{0.0, 0.2, 0.6}
\definecolor{forestgreen(traditional)}{rgb}{0.0, 0.5, 0.0}

\defcitealias{RVJ}{RVJ}

\defcitealias{MD17}{MD17}

\newcommand{\hf}{helium shell flash}
\newcommand{\hfs}{helium shell flashes}

\newcommand{\fgb}{first giant branch}
\newcommand{\agb}{asymptotic giant branch}

\newcommand{\fn}{FN~Sgr}
\newcommand{\koi}{KOI~3278}

\begin{document}

   \title{The formation of the magnetic symbiotic star FN Sgr}

   \titlerunning{Formation of the magnetic SySt FN Sgr}

   \author{Diogo Belloni\inst{1}
           \and
           Joanna Mikołajewska\inst{2}
           \and
           Matthias R. Schreiber\inst{1,3}
          }

    \authorrunning{Belloni, Mikołajewska \& Schreiber}

   \institute{Departamento de F\'isica, Universidad T\'ecnica Federico Santa Mar\'ia, Av. España 1680, Valpara\'iso, Chile\\
              \email{diogobellonizorzi@gmail.com}
              \and
              Nicolaus Copernicus Astronomical Center, Polish Academy of Sciences, Bartycka 18, 00-716 Warsaw, Poland
              \and
              Millenium Nucleus for Planet Formation, Valpara{\'i}so, Chile
             }

   \date{Received...; accepted ...}

 
  \abstract
   {
   There are several symbiotic stars (e.g.,], BF Cyg, Z And, and FN Sgr) in which periodic signals of tens of minutes have been detected. These periods have been interpreted as the spin period of magnetic white dwarfs that accrete through a magnetic stream originating from a truncated accretion disc.
   }
   {
   To shed light on the origin of magnetic symbiotic stars, we investigated the system FN Sgr in detail. We searched for a reasonable formation pathway to explain its stellar and binary parameters including the magnetic field of the accreting white dwarf.
   }
   {
   We used the MESA code to carry out pre-CE and post-CE binary evolution and determined the outcome of CE evolution assuming the energy formalism. For the origin and evolution of the white dwarf magnetic field, we adopted the crystallization scenario.
   }
   {
   We found that FN Sgr can be explained as follows. First, a non-magnetic white dwarf is formed through CE evolution. Later, during post-CE evolution, the white dwarf starts to crystallize and a weak magnetic field is generated. After a few hundred million years, the magnetic field penetrates the white dwarf surface and becomes detectable. Meanwhile, its companion evolves and becomes an evolved red giant. Subsequently, the white dwarf accretes part of the angular momentum from the red giant stellar winds. As a result, the white dwarf spin period decreases and its magnetic field reaches super-equipartition, getting amplified due to a rotation- and crystallization-driven dynamo. The binary then evolves into a symbiotic star, with a magnetic white dwarf accreting from an evolved red giant through atmospheric Roche-lobe overflow.
   }
   {
   We conclude that the rotation- and crystallization-driven dynamo scenario, or any age-dependent scenario, can explain the origin of magnetic symbiotic stars reasonably well. This adds another piece to the pile of evidence supporting this scenario. If our formation channel is correct, our findings suggest that white dwarfs in most symbiotic stars formed through CE evolution might be magnetic, provided that the red giant has spent ${\gtrsim3}$~Gyr as a main-sequence star.
   }

   \keywords{
             methods: numerical --
             stars: binaries: symbiotic --
             stars: evolution --
             stars: individual: FN Sgr --
             stars: magnetic field --
             white dwarfs
            }

   \maketitle
%


\section{Introduction}
\label{introduction}

Close white dwarf (WD) binaries are important for modern astrophysics for several reasons.
Most importantly, close WD binary stars are the progenitors of supernova Ia explosions \citep[e.g.,][]{webbink84-1,Maeda_2016,Liu_2023} and will be the dominant sources of low-frequency gravitational wave radiation to be detected with the Laser Interferometer Space Antenna \citep[][]{Kupfer_2018,Lamberts_2019,Scaringi_2023,AmaroSeoane_2023}.
Developing consistent models for the formation and evolution of close WD binaries is of fundamental importance in this context \citep[see][for a recent review]{Belloni_2023}.

Two recent discoveries have changed our understanding of WD binary formation and evolution.
First, the late appearance of magnetic fields in WDs \citep{Bagnulo_2021,Parsons_2021} can significantly alter their evolution \citep{Schreiber_2021a}.
Second, detailed simulations of the late AGB evolution are required to understand the population of observed post-common-envelope (CE) binaries with long orbital periods \citep{Belloni_2024b,Belloni_2024c}.
Symbiotic stars (SySts) offer the potential to further constrain both these processes.

SySts are interacting binaries in which, in most cases, a WD accretes matter from an evolved red giant (RG) donor, which can be a \fgb~(FGB) star or an \agb~(AGB) star \citep[e.g.,][]{Mikolajewska_2012,Munari_2019,Ilkiewicz_2022}, 
In SySts mass transfer is either driven by stellar winds or by atmospheric Roche-lobe overflow. In many cases, an accretion disc around the WD is formed due to conservation of angular momentum.
SySts seem to be absent in globular clusters \citep{Belloni_2020c} but are found in relatively large numbers in the field of the Milky Way and other galaxies \citep[e.g.,][]{Kniazev_2009,Mikolajewska_2014b,Mikolajewska_2017,Ilkiewicz_2018,Merc_2019,Merc_2020,Akras_2021,Merc_2021,Munari_2021,Merc_2022,Munari_2022,De_2022,Petit_2023,Akras_2023,Jia_2023}. 
Their orbital periods are typically between a few hundred days and a few thousand days, but it can also be tens of years if the RG is a Mira-like variable. 
This implies that SySts are likely the descendants from binaries with orbital periods typically exceeding a few hundred days consisting of a WD with a main-sequence star companion of mass ${\gtrsim1}$~\Msun.
We have recently shown that the few known potential progenitors of SySts can form through CE evolution if the detailed evolution of the WD progenitor on the AGB is taken into account \citep{Belloni_2024b,Belloni_2024c}.
As a logical next step we here start to investigate whether the same holds for SySts.

Recent indirect evidence indicates that some SySts contain strongly magnetic WDs.
Analysing the formation history of SySts can therefore provide additional constraints on the generation of WD magnetic fields. 
\citet{Mikolajewski_1990} observed a rapid coherent oscillation of about $500$~s in CH~Cyg and proposed that it could be explained as the rotation of a strongly magnetized WD (${\sim10}$~MG), that is, the observed flickering is a consequence from the interaction of the WD magnetosphere with the RG wind.
Other SySts with suspected magnetic WDs include Z~And which shows a stable oscillation with a $28$-min period that has been attributed to the spin period of an accreting magnetic WD \citep{Sokoloski_1999}.
Interestingly, \citet{Merc_2024} recently carried out an analysis of TESS light curves and found that this period has decreased $80$~s over a time span of $25$~yr, indicating that the WD has spun up.

Similarly, \citet{Formiggini_2009} discovered a coherent periodic signal of $108$~min in BF~Cyg and suggested that this signal originates from two hotspots on or near the surface of the WD due to magnetically-funnelled accretion.
In addition, \citet{Toma_2016} found a modulation on a period of $31.4$~min in the light curve of Hen~2–357, which would also be consistent with the rotation period of a magnetic WD.
Moreover, \citet{Magdolen_2023} found a periodic signal of $11.3$~min in the Kepler light curve of \fn~and suggested that this is the spin period of a magnetic WD accreting through a magnetic stream.
Most recently, \citet{Merc_2024} investigated periodic signals observed in the TESS light curves of some SySts.
They detected a signal in AE~Cir with a period of $27.3$~min and observed variability with a period of $66.6$~min in CI~Cyg, although the latter may not be related to the SySt (i.e., could be associated with a background source) due to the TESS very large pixels.
\citet{Merc_2024} also speculated that these two SySts could host magnetic WDs with rotation periods being revealed in the TESS data.
In all of the above listed systems but CH Cyg ellipsoidal variations are observed, which means that their RGs are close to filling their Roche lobes (i.e., ongoing atmospheric Roche lobe overflow) and their WDs accrete matter at rates ranging from a few ${10^{-8}}$ to a few ${10^{-7}}$~\Msunyr.

Among the several scenarios that have been proposed so far to explain the origin of magnetism in WDs, the crystallization scenario is currently the most promising one.
In brief, the idea of the rotation- and crystallization-driven dynamo in WDs can be described as follows.
After a carbon–oxygen WD is formed, it begins to age and cool down and when its temperature is sufficiently low, its inner region starts to crystallize and the crystallized mass fraction grows with time.
Crystallization leads to a solid oxygen rich core surrounded by a liquid carbon rich convective mantle, a configuration that can maintain a magnetic dynamo, similar to those assumed to generate the magnetic fields of planets, proto-stars and M dwarfs \citep{Christensen_2009}.
The idea that the onset of crystallization in WDs generates convection which then might drive a dynamo was first proposed by \citet[][]{Isern_2017} as an explanation for weak magnetic fields (${\sim0.1}$\,MG) but only gained considerable attention when \citet{Schreiber_2021a} showed that several problems of WD binary evolution can be solved if the dynamo is able to produce strong magnetic fields ${>1}$\,MG \citep[e.g.,][]{Schreiber_2021b,Belloni_2021,Bagnulo_2021,Schreiber_2022,Bagnulo_2022,Ginzburg_2022,Camisassa_2022,Saumon_2022,Fuentes_2023,Schreiber_2023,Caron_2023,Amorim_2023A,Pelisoli_2023,Pelisoli_2024,Blatman_2024,Hernandez_2024,CastroTapia_2024,Fuentes_2024}.

We here investigated whether we can explain the origin of magnetism in magnetic SySts with the rotation- and crystallization-driven dynamo. We focused our efforts on \fn, as this object is the SySt with the strongest evidence for hosting a magnetic WD. 
Its orbital period of $568.3$~d was derived by \citet{Brandi_2005}, based on long-term photometric data over 30 years.
The donor star in \fn~is an M\,5-type RG of mass ${\sim1.5}$~\Msun~and the accreting WD has a mass of ${\sim0.7}$~\Msun.
We carried out binary models and found that the properties of \fn~can be explained reasonably well through CE if the detailed evolution of the TP-AGB phase is taken in to account.
Less accurate evolutionary models such as BSE \citep{Hurley_2002} fail to reproduce the characteristics of the system.
The origin of the magnetic field of its WD is consistent with the crystallization dynamo scenario.

\begin{table}
\caption{Predicted and observed parameters of \fn. All observed parameters have been recalculated using the observational data from \citet{Brandi_2005}, except the orbital period which is from \citet{Magdolen_2023}.}
\label{Tab:Parameters}
\centering
\setlength\tabcolsep{3pt} 
\renewcommand{\arraystretch}{1.55} 
\begin{tabular}{lcc}
\hline\hline
Parameter & Observed & Predicted \\
\hline
inclination ($^{\rm o}$)                & $\gtrsim78$    &         \\
orbital period (d)                      & $567.3\pm0.3$  & $567.31$   \\
separation (\Rsun)                      & $358\pm19$     & $359.52$   \\
mass transfer rate ($10^{-8}$~\Msunyr)  & $2.6-5.2$      & $4.44$ \\
RG radius (\Rsun)                       & $150^{+15}_{-15}-161^{+13}_{-9}$   & $145.96$ \\
RG mass (\Msun)                         & $1.33\pm0.24$  & $1.331$ \\
RG Roche-lobe filling factor            & $\gtrsim 0.90$ & $0.906$ \\
WD mass (\Msun)                         & $0.60\pm0.09$  & $0.606$ \\
WD age (Gyr)                            &                & $2.883$ \\
\hline
\end{tabular}
\end{table}

\section{Revision of the parameters of FN~Sgr}
\label{Observation}

\citet{Magdolen_2023} have recently refined the orbital period of \fn~to ${567.3\pm0.3}$~d by combining photometric data from \citet{Brandi_2005} and \citet{Gromadzki_2013} with new data obtained in 2004-2022.
The new observations also revealed ellipsoidal variations,  particularly evident in their quiescent $I$-band light curve, which confirms the suggestion by \citet{Brandi_2005} that the RG fills or nearly fills its Roche lobe.

We have recalculated the orbital parameters using the radial velocity measurements from \citet{Brandi_2005} and adopting the revised orbital period from \citet{Magdolen_2023}.
The circular orbit solutions for the RG and the cF-type blue
absorption features result in a mass ratio of ${q=M_{\rm RG}/M_{\rm WD}=2.29\pm0.25}$, component masses of ${M_{\rm RG}\sin^3i=1.31\pm0.24}$~\Msun~and ${M_{WD}\sin^3i=0.57\pm0.09}$~\Msun, 
and a binary separation ${A\sin i=357\pm19}$~\Rsun, where $i$ is the binary inclination.

\citet{Brandi_2005} also derived the RG radius ${R_{\rm RG}/A=0.42\pm0.02}$ from the analysis of the shape and duration of the well-defined eclipses during the large 1996-2001 outburst.
However, if the system inclination is less than $90^{\rm o}$, this radius must be larger, with the maximum value corresponding to the Roche lobe radius which for our mass ratio is ${R_{\rm RL}=0.450\pm0.002}$.
Substituting the maximum ${R_{\rm RL}=0.452}$ and the minimum `eclipse' radius ${R_{\rm RG}/A=0.40}$ in eq.~2 in \citet{Brandi_2005}, we estimate ${i\geq78}$.
Finally, the resulting radius, that is ${\sim150-160}$~\Rsun, is consistent with a distance ${d\sim7.6-8.1}$~kpc, which is in a very good agreement with less accurate Gaia DR3 ${\sim8}$~kpc (but with a range ${6.33-10.15}$~kpc).

The hot WD luminosity varies between ${\sim2000-4000~(d/8\,{\rm kpc})^2}$~\Lsun\ \citep[][table 5]{Brandi_2005}. Moreover, the eclipse depth at quiescence remains more or less the same over the last 50 years \citep[][fig.1]{Magdolen_2023} which suggests similar luminosity of the eclipsed WD component as the values derived from spectroscopy.
If this luminosity is powered by nuclear burning, the mass accretion rate would be between ${\sim2.63\times10^{-8}}$ and ${\sim5.25\times10^{-8}}$~\Msunyr.
This is consistent with the predicted regime for stable nuclear burning \citep[e.g.,][]{Nomoto_2007} for the WD mass derived from observations (i.e., ${0.60\pm0.09}$~\Msun).
The parameters adopted for this study are summarized in Tab.~\ref{Tab:Parameters}.

\section{Binary models}
\label{Method}

\subsection{Pre- and post-CE evolution}

We used the version r15140 of the MESA code \citep[][]{Paxton2011,Paxton2013,Paxton2015,Paxton2018,Paxton2019,Jermyn2023} to calculate binary evolution from the zero-age main-sequence binary to the onset of the CE evolution, and from the moment just after CE evolution onward.
We describe here our assumptions for single star and binary evolution.
For reference, interested readers can access the MESA files that we made available\footnote{The files \texttt{run\_star\_extras.f90} and \texttt{run\_binary\_extras.f90} as well as the inlists needed for the simulations are available at \href{https://zenodo.org/records/10937460}{https://zenodo.org/records/10937460}}.
The MESA equation of state is a blend of the OPAL \citep{Rogers2002}, SCVH \citep{Saumon1995}, FreeEOS \citep{Irwin2004}, HELM \citep{Timmes2000}, PC \citep{Potekhin2010} and Skye \citep{Jermyn2021} equations of state.
Nuclear reaction rates are a combination of rates from NACRE \citep{Angulo1999}, JINA REACLIB \citep{Cyburt2010}, plus additional tabulated weak reaction rates \citep{Fuller1985,Oda1994,Langanke2000}.
Screening is included via the prescription of \citet{Chugunov2007} and thermal neutrino loss rates are from \citet{Itoh1996}.
Electron conduction opacities are from \citet{Cassisi2007} and radiative opacities are primarily from OPAL \citep{Iglesias1993,Iglesias1996}, with high-temperature Compton-scattering dominated regime calculated using the equations of \citet{Buchler1976}.

%
We adopted a metallicity of ${Z=0.02}$, which is consistent with the average metallicity ($[$Fe/H$]$\,$\sim-0.2$~dex) of RGs in Galactic disc and bulge SySts from the measurements of photospheric abundances of over 50 systems \citep[][]{Mikolajewska_2014a,Galan_2015,Galan_2016,Galan_2017,Galan_2023}.
We assumed the grey Eddington T(tau) relation to calculate the outer boundary conditions of the atmosphere, using a uniform opacity that is iterated to be consistent with the final surface temperature and pressure at the base of the atmosphere.
For the evolutionary phases with convective core, that is, core hydrogen and helium burning, we took into account exponential diffusive overshooting, assuming a smooth transition in the range ${1.2-2.0}$~\Msun~\citep[e.g.,][]{Anders_2023}.
We assumed that the extent of the overshoot region corresponds to ${0.016~H_{\rm p}}$ \citep[e.g.,][]{Schaller_1992,Freytag_1996,Herwig_2000}, with $H_{\rm p}$ being the pressure scale height at the convective boundary.
For the nuclear network, we assumed the auto-extended scheme, which automatically extend the net as needed.

%
We allowed the stars to lose mass through winds, adopting during pre-AGB evolution the \citet{Reimers_1975} prescription, setting the wind efficiency to $0.1$.
This value was motivated by observations of solar-metallicity open clusters, which suggest a low efficiency \citep{Miglio_2012,Handberg_2017}.
During AGB evolution, we adopted the prescription proposed by \citet{VW93}.
We included rotation in our calculations and we set the rotation of the zero-age main-sequence stars to $1$\% of their critical rotation rates.
%
%
We treated convective regions using the scheme by \citet{Henyey_1965} for the mixing-length theory, assuming that the mixing length is ${2\,H_{\rm p}}$ \citep[e.g.,][]{Joyce_2023}.
We also included mixing of angular momentum and rotationally induced mixing processes (Solberg-Hoiland, secular shear instability, Eddington-Sweet circulation, Goldreich-Schubert-Fricke and Spruit-Tayler dynamo), which in MESA are treated following \citet{Heger_2000} and \citet{Heger_2005}.

%
We allowed both the stars in the binary to eventually synchronize with the orbit, due to tidal interaction, except during post-CE evolution, in which we excluded the WD as its synchronization timescale is much longer than that of its companion.
%
%
The Roche-lobe radius of each star was computed using the fit of \citet{Eggleton1983}.
The mass transfer rates due to Roche-lobe overflow are determined following the prescription of \citet{Ritter1988}, in which the atmosphere of the star is filling the Roche lobe.
In this so-called atmospheric Roche-lobe overflow model, mass transfer occurs even when the star radius is smaller than the Roche-lobe radius.
Regarding wind accretion, we adopted the Bondi-Hoyle-Lyttleton prescription \citep{Hoyle_1939,Bondi_1944}.

%
We enforced that the main-sequence accretor ($M_{\rm a}$) cannot accrete the mass being transferred from the AGB donor ($M_{\rm d}$) at a rate higher than allowed by its thermal timescale ($\tau_{\rm KH,a}$).
We assumed that ${\dot{M}_{\rm a} = \eta\,\dot{M}_{\rm d}}$, where ${\eta=M_{\rm a}/\tau_{\rm KH,a}}$, if ${\dot{M}_{\rm d}>M_{\rm a}/\tau_{\rm KH,a}}$, or ${\eta=1}$, otherwise.
The non-accreted material is assumed to be lost from the vicinity of the main-sequence accretor as fast wind.
This critical rate is typically on the order of a few ${10^{-7}}$~\Msunyr.

%
For the post-CE evolution, we assumed that the WD is a point mass and evolved the model of the main-sequence companion saved at the onset of CE evolution.
The initial post-CE orbital period was calculated assuming the energy formalism as described in Sect.~\ref{CE}.
We further assumed circular orbits for the initial post-CE binaries. 
%
%
We enforced that the WD accretor can only accrete a part of the mass being transfer from its companion.
Depending on the accretion rate onto the WD, hydrogen shell burning could be stable, which results in an increase of its mass.
We implemented the critical accretion rate calculated by \citet[][]{Nomoto_2007}, above which WDs are thermally stable, that is, hydrogen burns steadily in a shell.
For accretion rates lower than this critical value, the WD undergoes nova eruptions, such that all of the accreted mass is assumed to be expelled from the binary.
However, we note that it is not very likely that exactly 100\% of the accreted material is ejected during a nova eruption \citep[e.g.,][]{Jose_2020}.
We further assumed that there is a maximum possible accretion rate \citep[][]{Nomoto_2007} such that WDs accreting at rates above it will burn stably at this maximum rate, and the remaining non-accreted matter will be piled up forming a red-giant-like envelope, which is assumed to be lost from the binary in the form of stellar-like winds.

We stopped both the pre-CE and post-CE evolution when the mass transfer rate reached ${10^{-2}}$~\Msunyr, to avoid halting the binary evolution earlier than the onset of the runaway mass transfer.
Finally, we adopted the default values for all other stellar and binary evolution parameters in MESA.

\subsection{CE evolution}
\label{CE}

For CE evolution we adopted the so-called energy formalism, in which the outcome of CE evolution is usually approximated by the balance between the change in the orbital energy and the envelope binding energy, given by \citep{Iben_Livio_1993}

\begin{equation}
E_{\rm bind} \ = \ 
\alpha_{\rm CE} \ \Delta E_{\rm orb} \ = \ - \ 
\alpha_{\rm CE} \
 \left( \, \frac{G\,M_{\rm d,c}\,M_{\rm a}}{2\,a_f}  \ - \ 
           \frac{G\,M_{\rm d,c}\,M_{\rm a}}{2\,a_i} \, \right) \ ,
\label{Eq:AlphaCE}
\end{equation}

\noindent
where $E_{\rm bind}$ is the envelope binding energy, $E_{\rm orb}$ is the orbital energy, $G$ is the gravitational constant, $M_{\rm d,c}$ is the core mass of the donor, $a_i$ is the semimajor axis at onset of the CE evolution, $a_f$ is the semimajor axis after CE ejection, and $\alpha_{\rm CE}$ is a parameter corresponding to the fraction of the difference in orbital energy (before and after CE evolution) that unbinds the envelope.

We calculated the binding energy by integrating over the star envelope from the helium core boundary (i.e., at radius at which the helium mass fraction is 0.1) to the surface of the star, that is,

\begin{equation}
E_{\rm bind} \ = \ 
- \int_{M_{\rm d,c}}^{M_{\rm d}} \frac{G \; m}{r(m)} \; {\rm d}m 
 \ + \  
\alpha_{\rm int} \int_{M_{\rm d,c}}^{M_{\rm d}} \varepsilon_{\rm int}(m) \; {\rm d}m \ ,
\label{Eq:AlphaINT}
\end{equation}

\noindent
where $r$ is the radius, $m$ is the mass, $\varepsilon_{\rm int}$ is the specific thermodynamic internal energy and $\alpha_{\rm int}$ is the fraction of the thermodynamic internal energy that are assumed to contribute to unbinding the envelope.
Our implementation to compute the thermodynamic internal energy is the same as in \citet{Belloni_2024c} and follows closely that by \citet{Hirai_2022}.

\begin{table*}
\centering
\caption{Evolution of a zero-age main-sequence binary towards \fn. For the pre-CE and post-CE evolution, we used the MESA code with the assumptions described in Sect.~\ref{Method}. For the CE evolution, we computed the post-CE orbital period using Eqs.~\ref{Eq:AlphaCE} and \ref{Eq:AlphaINT}, assuming ${\alpha_{\rm CE}=0.3}$ and ${\alpha_{\rm CE}\approx0.82}$. The terms $M_1$ and $M_2$ and Type$_1$ and Type$_2$ are the masses and stellar types$^{a}$ of the primary and secondary, respectively. $P_{\rm orb}$ is the orbital period and the last column corresponds to the event occurring to the binary at the given time in the first column. The row in which the binary has the present-day properties of \fn~is highlighted in boldface. For the WD evolution, we interpolated the evolutionary sequences calculated by \citet{Bedard_2020} to estimate the mass fraction of the crystallized matter. We assumed a magnetic field diffusion timescale of ${\sim600}$~Myr and that the amount of angular momentum accreted by the WD before the binary became a SySt was enough to spin up the WD towards the super-equipartition regime.}
\label{Tab:FormationChannel}
\setlength\tabcolsep{8pt} 
\renewcommand{\arraystretch}{1.25} 
\begin{tabular}{r c c c c r l}
\hline
\noalign{\smallskip}
 Time  &   $M_1$    &   $M_2$    & Type$_1$  & Type$_2$ & $P_{\rm orb}$ & Event\\
 (Myr) & (M$_\odot$)&(M$_\odot$) &           &          &  (days)        &      \\
\hline
\noalign{\smallskip}
     0.0000  &  2.200  &  1.360  & MS     & MS     &   2000.000  &  zero-age MS binary     \\
   840.5808  &  2.199  &  1.360  & SG     & MS     &   2001.400  &  change in primary type \\
   853.4929  &  2.199  &  1.360  & FGB    & MS     &   2001.459  &  change in primary type \\
   868.2086  &  2.198  &  1.360  & CHeB   & MS     &   2002.091  &  change in primary type \\
  1139.2842  &  2.196  &  1.360  & E-AGB  & MS     &   2004.652  &  change in primary type \\
  1156.3479  &  2.196  &  1.360  & TP-AGB & MS     &   1997.691  &  change in primary type \\
  1158.3386  &  2.116  &  1.360  & TP-AGB & MS     &   1785.599  &  onset of CE evolution $\left({\alpha_{\rm CE}=0.3}~{\rm and}~{\alpha_{\rm int}\approx0.82}\right)$\\
  1158.3386  &  0.606  &  1.360  & WD     & MS     &    630.000  &  end of CE evolution \\
  3096.4561  &  0.606  &  1.360  & WD     & SG     &    630.259  &  change in secondary type \\
  3349.8662  &  0.606  &  1.360  & WD     & SG     &    630.325  &  WD starts to crystallize\\
  3349.8662  &  0.606  &  1.360  & WD     & SG     &    630.325  &  WD weak magnetic field is generated deep in the core\\
  3624.6724  &  0.606  &  1.360  & WD     & FGB    &    630.401  &  change in secondary type \\
  3949.8662  &  0.606  &  1.359  & WD     & FGB    &    630.713  &  WD magnetic field penetrates the surface \\
  4041.4845  &  0.606  &  1.335  & WD     & FGB    &    578.405  &  WD is spun up (accreted enough angular momentum) \\
  4041.4845  &  0.606  &  1.335  & WD     & FGB    &    578.405  &  WD develops a super-equipartition strong magnetic field \\
  4041.7263  &  0.606  &  1.332  & WD     & FGB    &    568.701  &  onset of SySt phase (RL filling factor ${\approx0.90}$) \\
\textbf{4041.7403}  &  \textbf{0.606}  &  \textbf{1.331}  & \textbf{WD}     & \textbf{FGB} &   \textbf{567.310}  &  \textbf{binary looks like \fn} \\
  4041.7882  &  0.611  &  1.220  & WD     & FGB    &    497.086  &  onset of CE evolution $\left({\alpha_{\rm CE}=0.3}~{\rm and}~{\alpha_{\rm int}=1.0}\right)$\\
  4041.7882  &  0.611  &  0.452  & WD     & WD     &      0.954  &  end of CE evolution \\
\noalign{\smallskip}
\hline
\end{tabular}
\tablefoot{\tablefoottext{a}{Abbreviations:
MS~(main~sequence~star),
SG~(subgiant~star),
FGB~(first giant branch star),
CHeB~(core helium burning),
E-AGB~(early asymptotic giant branch star),
TP-AGB~(thermally-pulsing~asymptotic~giant~branch~star),
WD~(white~dwarf),
RLOF~(Roche~lobe~overflow),
CE~(common~envelope),
SySt~(symbiotic star).}}
\end{table*}

\begin{figure*}
\begin{center}
\includegraphics[width=0.99\linewidth]{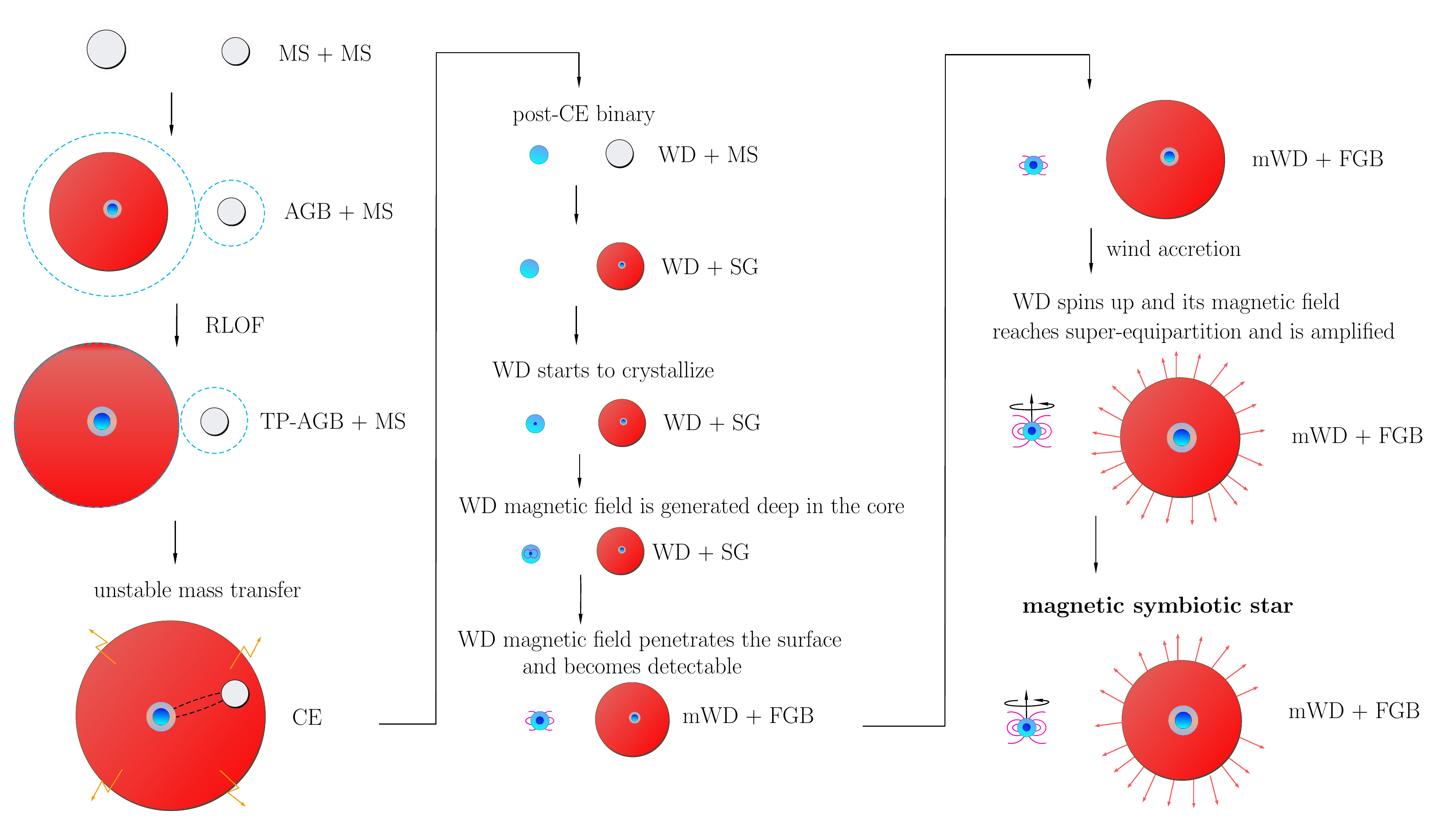}
\end{center}
\caption{Schematic illustrating the formation channel of \fn. A non-magnetic WD is first formed through CE evolution when the more massive star in the binary fills its Roche lobe as a TP-AGB star. The companion of the WD in the resulting post-CE binary evolves and becomes a sub-giant. Meanwhile, the WD ages and begins to crystallize, which creates the conditions for a weak magnetic field to be generated. The magnetic field initially remains deep inside the core but eventually, after a few hundred million years, diffuses and penetrates the surface, becoming detectable. In the meantime, the sub-giant star evolves, becomes an evolved FGB star, and starts to transfer some mass and angular momentum through stellar winds to the WD. As a result of the angular momentum accretion, the WD spin period decreases to minutes, and its magnetic field reaches the super-equipartition regime and is amplified, according to the rotation- and crystallization-driven dynamo. Subsequently, the binary evolves to a SySt, hosting a magnetic WD that is accreting matter through atmospheric Roche-lobe overflow.}
\label{Fig:Schematic}
\end{figure*}

\section{Toward a formation pathway for FN~Sgr}
\label{GRID}

We looked for a binary model to explain the properties of \fn~by running different grids of models.
We first searched for the range of the initial post-CE binary properties that during post-CE evolution results in a binary with the properties of \fn.
For that, we set the initial post-CE WD mass to $0.6$~\Msun~and varied the initial post-CE main-sequence mass (from $1.30$ to $1.40$~\Msun, in steps of $0.01$~\Msun) and orbital period (from $500$ to $700$~d, in steps of $10$~d).
We approximated the WD with a point mass and the assumed circular orbits.
We found that a binary with an initial post-CE main-sequence mass of ${\sim1.36}$~\Msun~and orbital period of ${\sim630}$~d evolves to a binary with properties comparable to \fn.

The next step was to search for the properties of the zero-age main-sequence binary such that, at the onset of the CE evolution, the hydrogen-free core of the RG donor, which is the future WD, is ${\sim0.6}$~\Msun, and the mass of the main-sequence companion is ${\sim1.36}$~\Msun.
In \citet{Belloni_2024b} and \citet{Belloni_2024c} we proposed two different pathways to explain long-period post-CE binaries.
We first proposed that the observed long-period post-CE binaries harbouring massive WDs (${\gtrsim1.2}$~\Msun) and AFGK-type main-sequence stars can be explained if the WD progenitor filled its Roche lobe when it was a highly evolved thermally-pulsing AGB (TP-AGB) star \citep{Belloni_2024b}.
This causes the onset of the CE evolution to take place at orbital periods of thousand days and makes the CE loosely bound requiring less orbital energy to unbind it, resulting in post-CE orbital periods as long as several hundred days.
Subsequently, we investigated the self-lensing binary \koi, which is a long-period post-CE binary (${\approx88}$~d) with a low-mass carbon-oxygen WD ($0.5250^{+0.0082}_{-0.0089}$~\Msun) and a G-type main-sequence stars ($0.911^{+0.023}_{-0.026}$~\Msun).
We showed that, without invoking energy sources other than gravitational and thermodynamic internal, binaries like \koi, that is, with long orbital period and containing low-mass carbon-oxygen WDs, can only be formed if the WD progenitor fills its Roche lobe at a \hf~in the beginning of the TP-AGB evolution, as otherwise either the WD mass becomes too high, or the mass transfer does not become dynamically unstable, or the resulting post-CE orbital period will be too short \citep{Belloni_2024c}.

We first investigated whether the first scenario outlined above could also lead to long-period post-CE binaries with typical WD masses as required for \fn.
We determined that this pathway may not be effective due to the fact that the WD progenitor structure at the onset of the CE evolution is such that mass transfer is always dynamically stable, regardless of whether it fills its Roche lobe during \hfs~or between pulses.

As a next step we inspected whether a less evolved TP-AGB star could be the progenitor of the WD in \fn, that is, a TP-AGB star that underwent considerable core mass growth but almost negligible mass loss through stellar winds.
We indeed found that this configuration leads to dynamically unstable mass transfer.
A more thorough investigation of the stability of mass transfer from TP-AGB donors and the impact of its structure, the growth of its core mass, and mass loss through winds will be presented in another paper.

Bearing these results in mind, to find a reasonable binary model for \fn, we assumed in our grid of pre-CE models that the WD progenitor originates from a zero-age main-sequence star that does not lose a substantial amount of mass during the TP-AGB evolution, but at the same time fills its Roche lobe after several thermal pulses, allowing in turn its core to grow in mass to ${\sim0.6}$~\Msun, before filling its Roche lobe.
Furthermore, we assumed that the initial orbit was circular and, unlike the previous step regarding the post-CE evolution, we also evolved the companion of the WD progenitor during pre-CE evolution.
This was done because we used the model of the main-sequence companion obtained at the onset of the CE evolution to feed the initial post-CE evolution for simulating the post-CE evolution.

\begin{figure}
\begin{center}
\includegraphics[width=0.99\linewidth]{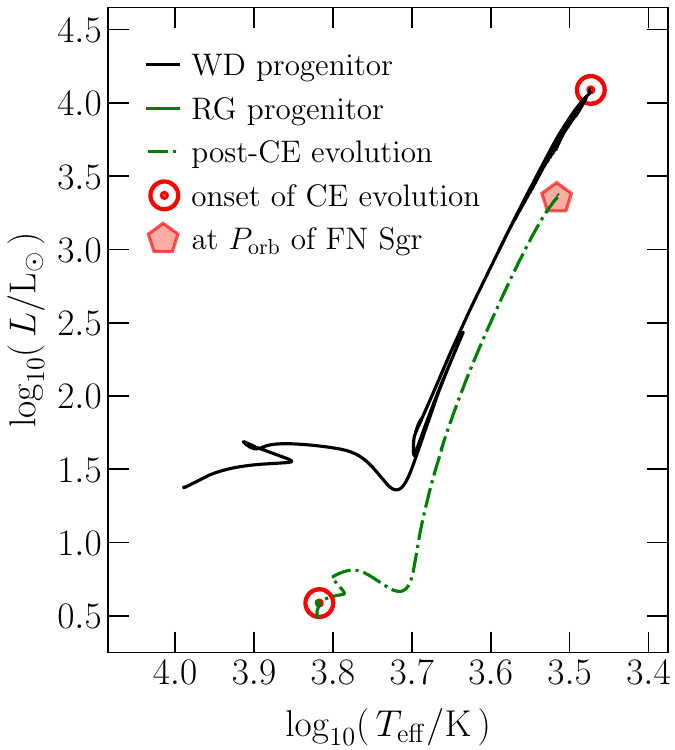}
\end{center}
\caption{Hertzsprung-Russell diagram of both stars in the binary. The black line corresponds to the WD progenitor, that is, the initially more massive star, up to the onset of CE evolution (red solar symbol). During pre-CE evolution, the initially less massive star, which is the RG progenitor (represented by the green lines) evolved only slightly. After CE evolution (dot-dashed line), the companion evolved, became an evolved RG and quickly afterward its atmosphere filled its Roche lobe. At this moment the binary became a SySt and, shortly after that (red pentagon), its properties were similar to those of \fn.}
\label{Fig:HRdiagram}
\end{figure}

\begin{figure}
\begin{center}
\includegraphics[width=0.94\linewidth]{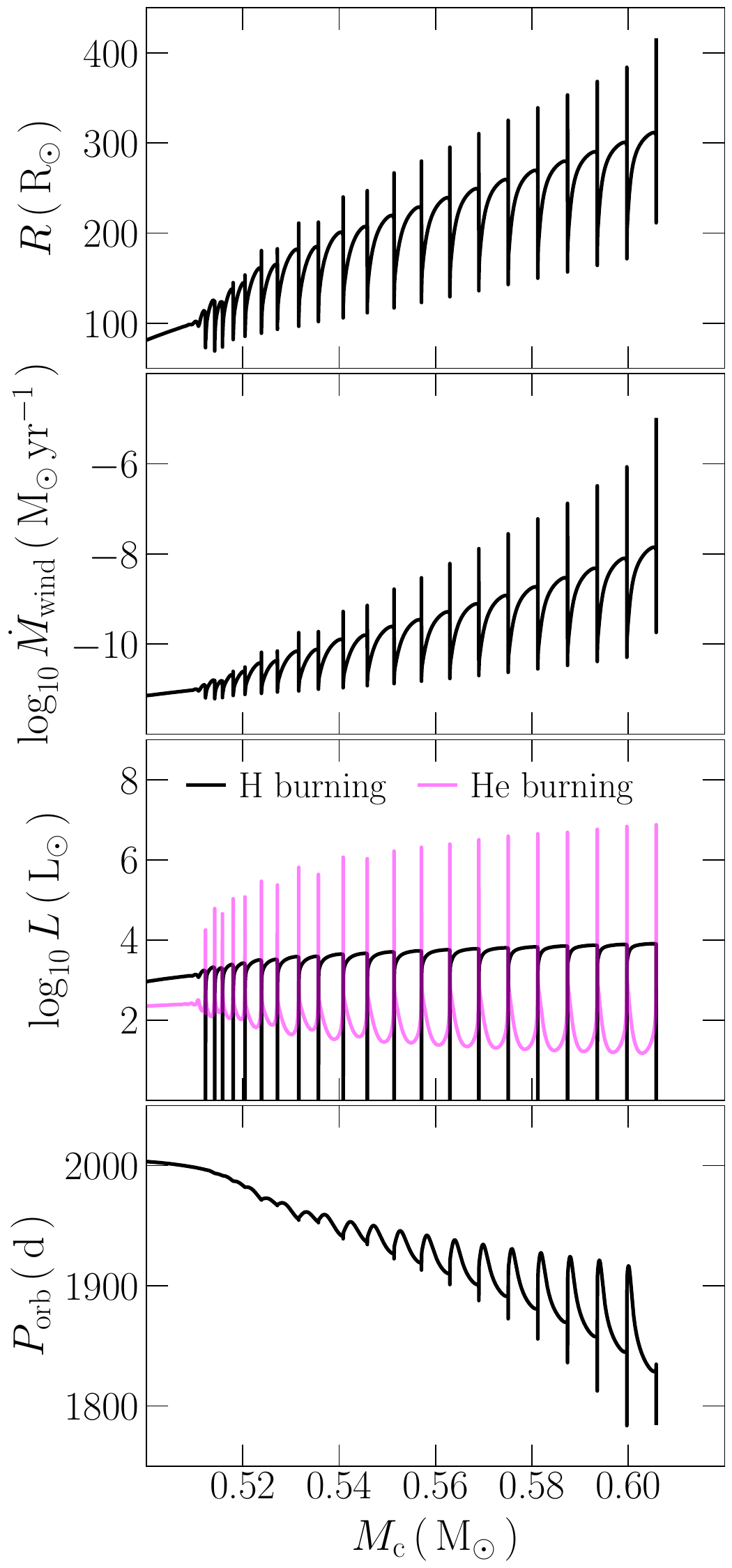}
\end{center}
\caption{Pre-CE evolution of the WD progenitor covering the TP-AGB phase up to the onset of the CE evolution, which takes place at the last \hf~shown. We show the evolution of the orbital period ($P_{\rm orb}$), the hydrogen and helium burning luminosities ($L$), the wind mass loss rate ($\dot{M}_{\rm wind}$), and the radius, as a function of the mass of the hydrogen-free core. Although several thermal pulses occurred before the WD progenitor filled its Roche lobe, leading to substantial core mass growth, the TP-AGB star is not yet highly evolved, as evidenced by its radius and wind mass loss rate. This causes the mass transfer to be dynamically unstable which generates CE evolution.}
\label{Fig:preCE}
\end{figure}

\section{Formation pathway for FN Sgr}
\label{FN}

From the exercise in the previous section, we found that a binary with zero-age main-sequence masses of ${\sim2.2}$ and ${\sim1.36}$~\Msun~and orbital period of ${\sim2000}$~d, evolves to the required initial post-CE binary to explain \fn.
In what follows we will discuss in more detail how a zero-age main-sequence binary with these properties evolves to a binary having the properties of \fn.
The sequence of events is given in Tab.~\ref{Tab:FormationChannel}, while we illustrate our scenario in Fig.~\ref{Fig:Schematic}.

\subsection{Pre-CE and CE evolution}
\label{FNpreCE}

Initially, the more massive star evolves and becomes a TP-AGB star after ${\approx1.15}$~Gyr.
The onset of CE evolution takes place during a \hf, after several thermal pulses, when the TP-AGB donor star has a mass ${\approx2.1}$~\Msun~and its hydrogen-free core has a mass of ${\approx0.6}$~\Msun.
Throughout pre-CE evolution, the mass of the main-sequence star remains practically the same as the mass loss from the WD progenitor is not sufficiently high, which makes wind accretion inefficient.

We show in Fig.~\ref{Fig:HRdiagram} the evolution of both stars in a Hertzsprung-Russell diagram.
Additionally, we show in Fig.~\ref{Fig:preCE} the evolution of the the orbital period, the hydrogen and the helium burning luminosities, the wind mass loss rate, and the radius, as a function of the TP-AGB core mass. 
At the beginning of the TP-AGB evolution, the core mass is ${\approx0.51}$~\Msun.
After 20 thermal pulses, the core mass increased to ${\approx0.60}$~\Msun, while the orbital period decreased to ${\sim1800}$~d due to tidal interactions.
In addition, the wind mass loss rate varied from ${\sim10^{-11}}$ to ${\sim10^{-8}}$~\Msunyr, while the radius increased from ${\sim100}$ to ${\sim300}$~\Rsun.
The onset of the CE evolution took place at the next pulse.

\begin{figure}
\begin{center}
\includegraphics[width=0.99\linewidth]{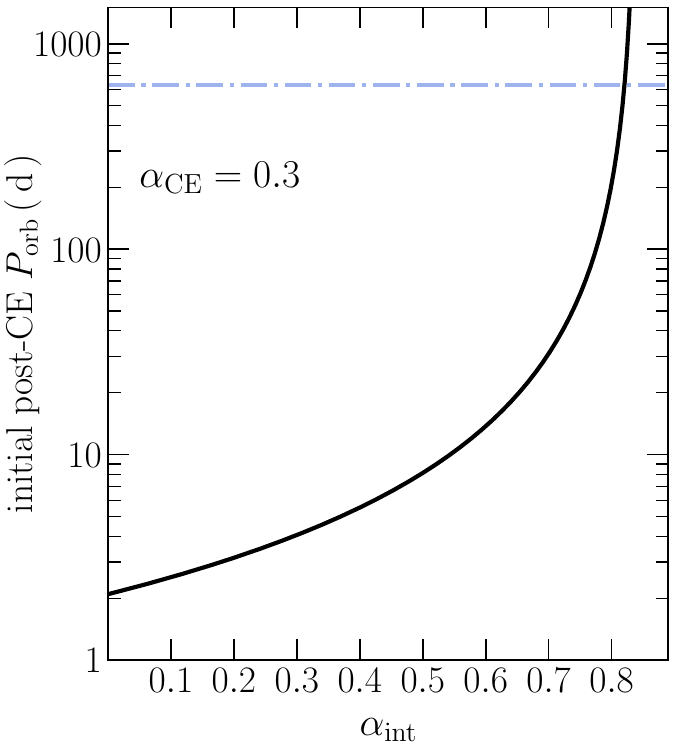}
\end{center}
\caption{Orbital period ($P_{\rm orb}$) just after the CE evolution as a function of the fraction of the thermodynamic internal energy ($\alpha_{\rm int}$) used to calculate the CE binding energy (Eq.~\ref{Eq:AlphaINT}), assuming ${\alpha_{\rm CE}=0.3}$. The horizontal line indicates the initial post-CE orbital period that is required to explain \fn, which is obtained by assuming ${\alpha_{\rm int}\approx0.82}$.}
\label{Fig:CE}
\end{figure}

The CE binding energy was computed with Eq.~\ref{Eq:AlphaINT}.
For ${\alpha_{\rm int}=0}$, the binding energy is simply the gravitational energy, which corresponds to ${-5.39\times10^{46}}$~erg.
The thermal energy, which is part of the thermodynamic energy, is ${2.83\times10^{46}}$~erg, while the integral thermodynamic internal energy is ${6.48\times10^{46}}$~erg.
We can extract two important pieces of information from these values.
First, when either only gravitational energy or gravitational energy plus thermal energy are assumed to compute the CE binding energy, only short-period post-CE binaries are expected.
Second, it is clear that the only a fraction of ${\alpha_{\rm int}\lesssim83}$\% of thermodynamic internal energy can contribute to CE binding energy as otherwise the binding energy would be positive.
We show in Fig.~\ref{Fig:CE} how the resulting orbital period of the post-CE binary depends on $\alpha_{\rm int}$, assuming ${\alpha_{\rm CE}=0.3}$ \citep{Zorotovic_2010}.
The required orbital period to explain \fn~is obtained when ${\alpha_{\rm int}\approx0.82}$.

\begin{figure}
\begin{center}
\includegraphics[width=0.94\linewidth]{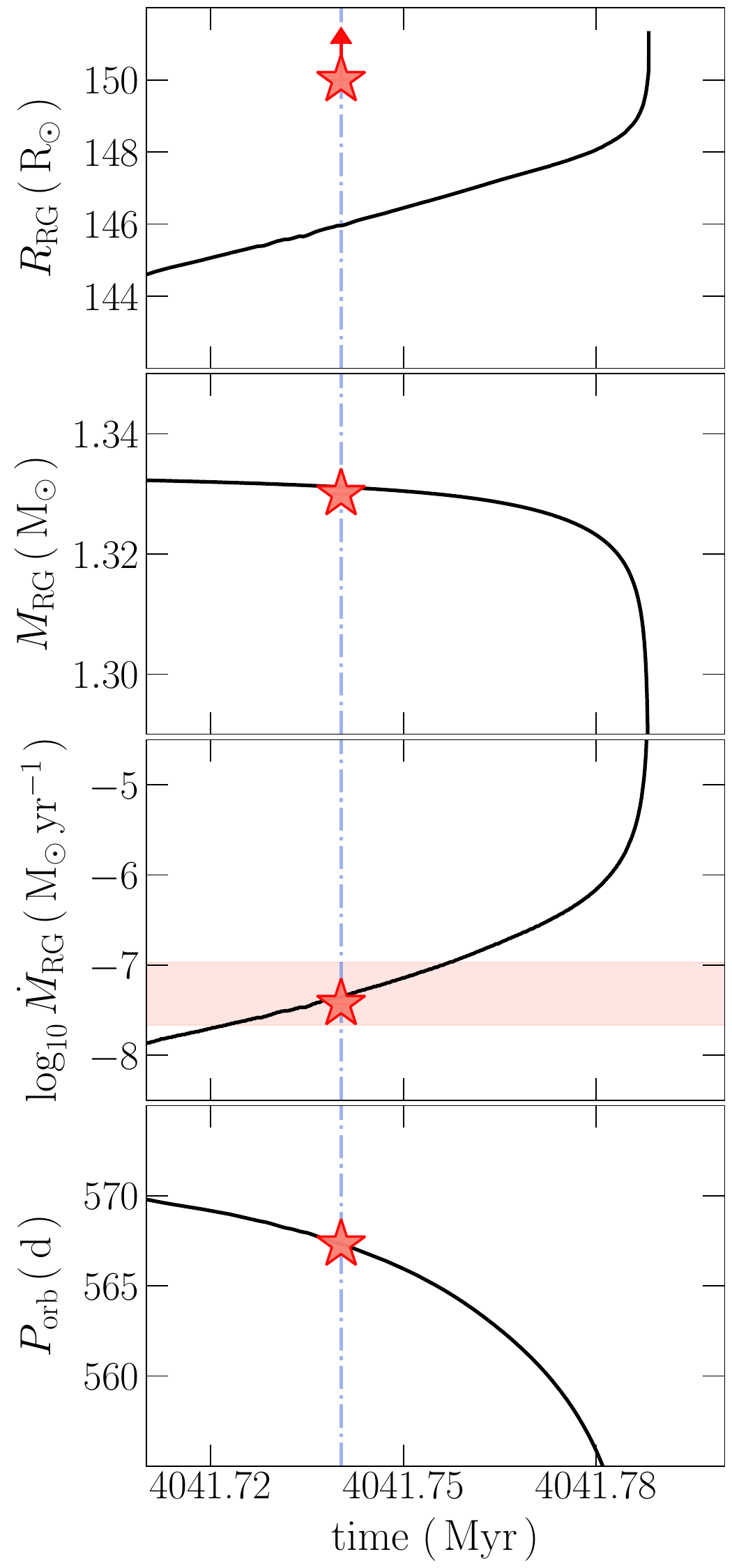}
\end{center}
\caption{Post-CE evolution with time of the orbital period ($P_{\rm orb}$), the mass transfer rate due to atmospheric Roche-lobe overflow ($\dot{M}_{\rm RG}$), RG mass ($M_{\rm RG}$) and RG radius ($R_{\rm RG}$). The red star indicates the present-day properties of \fn, without error bars, the red rectangle the minimum and maximum mass transfer rate required to explain the WD luminosity as powered by nuclear burning, and the vertical blue line the time at which the binary has the same orbital period of \fn. Our model reproduces reasonably well not only the RG mass, RG radius and the orbital period but also the mass transfer rate.}
\label{Fig:postCE}
\end{figure}

\subsection{Post-CE evolution}
\label{FNpostCE}

The resulting post-CE binary has an initial orbital period of $630$~d, it hosts a non-magnetic WD of mass $0.6058$~\Msun~and a main-sequence star of mass $1.36$~\Msun.
During the early post-CE binary evolution, the companion evolves and becomes a sub-giant.
Meanwhile, the WD cools and begins to crystallize at an age of ${\approx2.191}$~Gyr, after a total evolutionary time of ${\approx3349}$~Myr.
At this moment, the crystallization driven dynamo can lead to the generation of a magnetic field.

As shown by \citet{Ginzburg_2022}, for typical WD spin periods, the magnetic energy density is expected to reach an equipartition with the kinetic energy.
Despite that, the generated magnetic field should be weak.
In addition, these authors also showed that the magnetic field remains initially deep inside the core of the WD and only after a diffusion timescale of several hundred Myr the magnetic field can penetrate the surface and become detectable.
We found that a diffusion timescale of at most ${\sim600}$~Myr is required to explain \fn~with the masses we assumed.
If this timescale is longer, then the WD in \fn~should not host a detectable magnetic field according to the crystallization scenario. 
A longer diffusion timescale would require the RG mass to be lower than we assume, which would translate to a longer evolutionary timescale and in turn allow the WD to cool down for longer.

After the WD magnetic field was generated, the WD companion kept evolving and became an FGB star at a time ${\approx3624}$~Myr.
We assumed that the WD magnetic field reached the surface at a time ${\approx3949}$~Myr, when it is ${\approx2.883}$~Gyr old, corresponding to a crystallized mass fraction of ${\approx0.18}$.
Meanwhile, the WD companion became an evolved FGB star, with a sufficiently high mass loss rate so that part of this mass is effectively transferred to the WD. 
Even though this rate is not high enough to prevent nova eruptions, that is, all accreted mass is ejected in such events, some of the accreted angular momentum remains in the WD.
The amount of mass transferred to the WD before the binary becomes a SySt is ${\sim10^{-3}}$~\Msun.
As shown by \citet{Schreiber_2021b}, this amount of accreted mass is enough to spin up the WD to periods of the order of minutes/hours.


As the WD became a fast rotator, the conditions for the overall magnetic energy to reach super-equipartition levels compared to the convective kinetic energy are satisfied \citep{Augustson_2016,Augustson_2019}.
This implies that the magnetic field can be amplified and in turn become much stronger than initially.
We assumed that the amplification timescale is comparable to the spin up timescale so that the magnetic field smoothly gets stronger as the WD spin period decreases.
Quickly afterward, the binary evolves to a symbiotic star with properties comparable to those of \fn, hosting a magnetic WD that is accreting matter from a Roche-lobe-filling RG.
During the SySt phase, the WD continues to accrete angular momentum and spin up to the observed spin period.

%
%

We show in Fig.~\ref{Fig:postCE} the post-CE evolution from around the moment the binary becomes a SySt up to the onset of the second CE evolution.
We included the evolution of the orbital period, mass transfer rate, RG mass and RG radius, as these properties can be directly compared with those derived from observations.
All predicted and observed values are in reasonably good agreement, as can be seen in Tab.~\ref{Tab:Parameters}.
%
%
In particular, to explain the observed WD luminosity, including its quiescent and outburst characteristics, the average mass transfer rate is expected to lie within the range where hydrogen is burnt steadily on the WD surface, without exceeding the critical rate above which the WD envelope expands to a giant size, which occasionally occurs and can be observed as an optical outburst.


We predict that \fn~will quickly evolve to a second episode of CE evolution (within ${\approx0.05}$~Myr).
This event will likely result in a post-CE double WD binary having a magnetic carbon-oxygen WD and a non-magnetic helium WD.
Assuming ${\alpha_{\rm CE}=0.3}$, the resulting orbital period will range from ${\sim5}$~h (for ${\alpha_{\rm int}=0.0}$) to ${\sim1}$~d (for ${\alpha_{\rm int}=1.0}$).


\section{Discussion}
\label{Discussion}

We have presented an evolutionary scenario for the magnetic SySt \fn.
Our model assumes that the magnetic field is generated by a magnetic dynamo driven by crystallization.
The formation of the binary consisting of a WD and a RG companion was calculated with MESA and required the progenitor of the WD to fill its Roche-lobe on the TP-AGB phase after several pulses.
In what follows we discuss our assumptions concerning the WD magnetic field generation, alternative scenarios for the origin of WD magnetic fields, as well as implications of our results for close binary star formation.

\subsection{Alternative models for the generations of magnetic fields}

Scenarios other than the crystallization we assumed here do not seem to be viable routes to explain the origin of the magnetic WD in \fn.
For instance, the CE dynamo scenario \citep[e.g.,][]{Tout_2008} can be easily ruled out, not only because of the several problems it faces \citep{Belloni_2020b} but mainly because the post-CE orbital period is too long so that only a negligible non-detectable magnetic field would be predicted according to this scenario.

Another possibility would be the double degenerate merger scenario \citep{Berro_2012}, which in this case requires that the system originated from a triple, in which the inner binary merged resulting in a magnetic WD.
This scenario is unlikely because the post-merger orbital period required to explain the observed orbital period of \fn~is in principle too short to be explained with triple evolution, which would result in a post-merger binary much wider than required.

The last possibility would be the fossil field scenario.
Albeit we cannot completely rule out this channel, it faces many problems, not only for single WDs but also for WDs in binaries.
For example, it is not clear how to explain with this scenario the late appearance of magnetic fields in volume-limited samples \citep{Bagnulo_2021} as well as the lack of progenitors of strongly magnetized WDs in close semi-detached binaries \citep{Parsons_2021,Schreiber_2021b}.

\subsection{The rotation- and crystallization-driven dynamo}


Inspired by \citet{Schreiber_2021a}, we have assumed that a phase in which the WD spins up is required for the generation of a strong and detectable magnetic field. More recent results indicate that the WD does need to be rotating as fast as we assumed (spin periods shorter than ${\sim1}$~min).
While observations show that magnetic WDs rotate on average faster than non-magnetic WDs, some strongly magnetized rotate with periods of the order of hours/days \citep{Hernandez_2024}.

This is roughly consistent with the theoretical calculations performed by \citet{Ginzburg_2022}, which suggest that spin periods of the order of one hour can lead to magnetic field strengths ${\gtrsim1}$~MG.
However, \citet{Ginzburg_2022} also argued that for typical WD spin periods, only weak magnetic fields (${\lesssim0.1}$~MG) are expected, which indicates that either the strong magnetic fields in these slow rotators originate from a different mechanism or the theory to estimate the magnetic field strength in the crystallization scenario is not fully understood.

Taking into account all these observational and theoretical results, we proposed a scenario in which first a weak magnetic field is generated in the core as soon as the WD starts to crystallize.
Subsequently, this magnetic field diffuses to the surface and, as the WD accretes angular momentum from its RG companion via wind accretion, the magnetic field gradually reaches super-equipartition and gets amplified and becomes detectable.
Our scenario relies on the assumption that the amplification timescale is short enough and comparable to the angular momentum accretion timescale, allowing the WD magnetic field to get stronger as the WD spin period decreases.
Further theoretical WD models addressing the specific issue of how the magnetic field evolves as the WD accretes matter and spins up are required to test this assumption.

\subsection{The diffusion timescale}


As argued by \citet{Ginzburg_2022}, the magnetic field is initially trapped in the convection zone, deep inside the carbon-oxygen core.
After several hundred Myr, it diffuses, penetrates the surface, and becomes detectable.
More recently, \citet{Blatman_2024} improved those calculations with MESA models by including phase separation and taking into account a more consistent WD structure.
They predicted that roughly half of the WD mass has to crystallize before the magnetic field emerges to the surface, leading to diffusion timescales of the order of a few Gyr.
If the magnetic field takes that long to reach the surface, our proposed scenario is very unlikely to be true.

However, \citet{Blatman_2024} could not explain the magnetic field appearance of `all' WDs in the volume-limited sample built by \citet{Bagnulo_2022}, especially those with a mass of ${\approx0.7}$~\Msun.
This remains true even if uncertainties in the phase diagram and in the nuclear reaction rates are taken into account.
Nevertheless, \citet{Blatman_2024} highlighted that there are other processes as well as uncertainties in the treatment of the internal structure of WDs that could explain the disagreement.
For instance, convective core overshooting during core helium burning may significantly alter the initial carbon-oxygen profiles before the onset of crystallization, affecting in turn the diffusion timescale.
Further calculations are required to provide stronger constraints on the diffusion timescales.

\begin{figure}
\begin{center}
\includegraphics[width=0.99\linewidth]{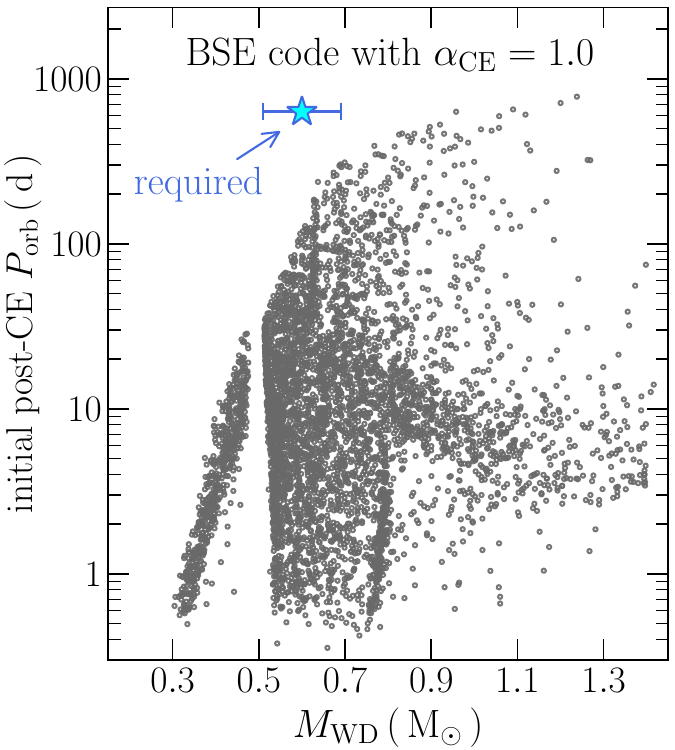}
\end{center}
\caption{Orbital period ($P_{\rm orb}$) just after the CE evolution as a function of the WD mass ($M_{\rm WD}$) from the post-CE population synthesis carried out by \citet{Belloni_2024b}, assuming ${\alpha_{\rm CE}=1.0}$. We show only systems with companion masses of ${1.33\pm0.24}$~\Msun~to be consistent with observations. The blue star indicates the initial post-CE orbital period and WD mass that are required to explain \fn.}
\label{Fig:BSE}
\end{figure}

\subsection{Alternative evolutionary scenarios}


We shall notice that a scenario involving dynamically stable non-conservative mass transfer is unlikely for a couple of reasons.
First, a WD with a mass of ${\sim0.6}$~\Msun~originates from a star that is much more massive than the required RG mass at the end of the mass transfer (i.e., ${\sim1.3}$~\Msun).
This implies that mass transfer is very unlikely to have been stable, and the situation is even worse if the WD progenitor fills its Roche lobe at an earlier evolutionary stage than we propose, as the required mass of this star would be higher than what we assumed (i.e., ${2.2}$~\Msun) to produce a WD of $0.6$~\Msun.
Second, for reasonable assumptions of orbital angular momentum loss during dynamically stable mass transfer, it is unlikely that the orbital period could decrease by a factor of ${\sim5}$ as required.
We therefore conclude that the only reasonable formation channel for \fn~is CE evolution, as discussed in Sect.~\ref{GRID}.

\citet{Belloni_2024b} have recently carried out post-CE binary population synthesis with the BSE code assuming in the CE energy budget only orbital, gravitational, and thermal energies.
We proposed that the systems harbouring massive WDs (${\gtrsim1.1}$~\Msun) with long orbital periods descended from highly evolved TP-AGB stars.
We show in Fig.~\ref{Fig:BSE} their results for systems with main-sequence star companions of masses ${1.33\pm0.24}$~\Msun~and ${\alpha_{\rm CE}=1.0}$.
It is quite clear from the figure that BSE cannot explain the post-CE binary properties that are required to explain \fn.
To reproduce \fn~with BSE, one has to assume that extra energy, such as recombination energy, contributes to the CE ejection.
There is an ongoing debate about whether recombination energy can affect CE ejection \citep[e.g.,][and references therein]{Soker_2003,webbink84-1,Ivanova_REVIEW,Zorotovic_2014,Nandez_2015,Ivanova_2015,Ivanova_2018,Soker_2018,Grichener_2018,R2020,K2020,LC2022,GB2022,Lau_2022,Belloni_2023,Ropke_2023,Chen_2024}

We found here that long-period post-CE binaries with less massive WDs cannot form in the way suggested by \citet{Belloni_2024b} because mass transfer tends to be dynamically stable, which results in an orbital period different from what is observed.
In addition, with our detailed TP-AGB modelling, we showed that no energy other than gravitational and thermodynamic internal is needed to explain \fn.
This has two important implications.
First, this serves as a clear example of how the simplified assumptions of (at least some) fast population synthesis codes can lead to misleading results.
Second, it highlights the significance of incorporating the detailed structure of TP-AGB donors in binary models to better understand the CE evolution and its outcomes.

\subsection{Implications for other systems}


While we only modelled evolutionary pathways (including the generation of the magnetic field) for the \fn, other potentially magnetic SySts seem to be consistent with having formed in the same way.
Performing detailed modelling dedicated to explain these objects, however, represents a rather futile exercise as long as several parameters of these SySts have not yet been reliably measured. 
For instance, for Z~And, accurate measurements of the WD and RG masses are not yet available. Similarly, for BF~Cyg only minimum masses have been published, that is ${1.8\pm0.6}$ and ${0.5\pm0.1}$~\Msun, for the RG and the WD, respectively \citep{Fekel_2001}.
Future observational efforts aiming at more accurately constraining these parameters would be useful to further test the results achieved in this work.
Providing these constraints, however, might be observationally challenging as these two systems are not eclipsing.


Finally, our scenario has implications beyond SySts as we predicted that the fate of \fn~will be a double WD binary composed of a magnetic carbon-oxygen WD orbiting a helium WD with an orbital period of ${\sim1}$~d.
The origin of magnetic double WD binaries has been addressed by \citet{Schreiber_2022}, who proposed a scenario involving a phase in which the binary was a SySt to explain the only known system  \citep[i.e., NLTT~12758][]{Kawka_2017}. 
Both results taken together suggest that SySts could be crucial for understanding the incidence of magnetism among close double WD binaries.

\section{Conclusions}\label{Conclusion}

We carried out binary evolution with the MESA code and investigated the origin of magnetic symbiotic stars.
We proposed that the future accreting WD is formed through CE evolution when the more massive star in the binary fills its Roche lobe as an asymptotic giant branch star that underwent substantial core mass growth but only negligible mass loss through stellar winds.
This leads to the formation of a non-magnetic WD paired with a main-sequence star.
During the early post-CE binary evolution, the WD cools, eventually starts to crystallize, and a weak magnetic field is generated, which initially remains deep inside the core.
After a few hundred million years, the magnetic field diffuses to the surface and becomes detectable.
Meanwhile, the main-sequence star evolves, becomes a first giant branch star, and eventually starts transferring part of the angular momentum from its stellar winds to the WD.
As a result of this angular momentum accretion, the WD spins up to periods of the order of minutes/hours and the magnetic field is amplified.
When the binary finally becomes a symbiotic star, it hosts an accreting magnetic WD in which matter is funnelled by its magnetic field to the regions near the magnetic poles.
Afterward, this magnetic symbiotic star will evolve to a close binary having a magnetic WD orbiting another WD, which suggests that symbiotic stars might be key to understanding the incidence of magnetism among close double WD binaries.

Our results support the idea that a crystallization driven dynamo is responsible for the generation of magnetic fields in WDs although theoretical issues, such as the diffusion timescale as well as the appropriate scaling law for determining the field strength still need to be solved. 
With respect to the formation of close binary stars, especially those containing a WD with an main sequence or RG companion, our simulations further show that recombination energy is not required during CE evolution as long as the TP-AGB evolution of the WD progenitor is properly modelled.

\begin{acknowledgements}

We would like to thank an anonymous referee for the comments and suggestions that helped to improve this manuscript.
We thank the Kavli Institute for Theoretical Physics (KITP) for hosting the program ``White Dwarfs as Probes of the Evolution of Planets, Stars, the Milky Way and the Expanding Universe''.
This research was supported in part by the National Science Foundation under Grant No. NSF PHY-1748958.
This research was partially supported by the Munich Institute for Astro-, Particle and BioPhysics (MIAPbP) which is funded by the Deutsche Forschungsgemeinschaft (DFG, German Research Foundation) under Germany's Excellence Strategy -- EXC--2094 -- 390783311.
DB acknowledges financial support from {FONDECYT} grant number {3220167}.
JM was supported by the Polish National Science Centre (NCN) grant 2023/48/Q/ST9/00138.
MRS was supported from {FONDECYT} grant number {1221059} and ANID, – Millennium Science Initiative Program – NCN19\_171.

\end{acknowledgements}

%
%

\bibliographystyle{aa} 
\bibliography{references} 

\end{document}